%% file: paper.tex
\documentclass[11pt,a4paper]{article}
\usepackage{jheppub}
\usepackage{graphicx}
\pdfoutput=1
\usepackage{url}
\usepackage{cancel}
\usepackage{caption}
\usepackage{feynman}
\usepackage{lscape}
\usepackage{bm}
\usepackage{hyperref}
\usepackage{amsmath}
\usepackage{amssymb}
\usepackage{listings}
\usepackage{enumitem}
\usepackage[toc, page]{appendix}
\usepackage{color}
\usepackage[section]{placeins}

\newcommand{\order}[1]{{\cal O}(#1)}

\def\as{\alpha_s}

\def\G0{{\bf \gamma^{(0)}_N}}
\def\G1{{\bf \gamma^{(1)}_N}}

\newcommand{\beq}{\begin{equation}}
\newcommand{\eeq}{\end{equation}}
\newcommand{\bea}{\begin{eqnarray}}
\newcommand{\eea}{\end{eqnarray}}
\newcommand{\li}[1]{ {\rm Li}_{#1}}
\newcommand{\mell}[1]{\mathcal M\left[{#1}\right]}

\def\gev{{\rm GeV}}

\title{Resummation effects in the bottom-quark fragmentation function}
\author[a]{Fabio Maltoni,}
\author[b]{Giovanni Ridolfi,}
\author[c]{Maria Ubiali,}
\author[d]{Marco Zaro}
\affiliation[a]{Centre for Cosmology, Particle Physics and Phenomenology (CP3),
Universit\'e catholique de Louvain, 1348 Louvain-la-Neuve, Belgium;
Dipartimento di Fisica e Astronomia, Universit\`a di Bologna and INFN,
Sezione di Bologna, via Irnerio 46, I-40126 Bologna, Italy}
\affiliation[b]{Dipartimento di Fisica, Universit\`a di Genova and INFN, Sezione di Genova, Via Dodecaneso 33, I-16146 Genova, Italy}
\affiliation[c]{DAMTP, University of Cambridge, Wilberforce Road, Cambridge, CB3 0WA, United Kingdom}
\affiliation[d]{Tif Lab, Dipartimento di Fisica, Universit\`a di Milano and INFN, Sezione di Milano, Via Celoria 16, I- 20133 Milano, Italy}
\emailAdd{fabio.maltoni@uclouvain.be}
\emailAdd{giovanni.ridolfi@ge.infn.it}
\emailAdd{M.Ubiali@damtp.cam.ac.uk}
\emailAdd{marco.zaro@mi.infn.it}

\abstract{
We compute the perturbative component of the fragmentation function of
the $b$ quark to the best of the present theoretical knowledge. The
fixed-order calculation to order $\as^2$ of the fragmentation function
at the initial scale is matched with soft-emission logarithm resummation to
next-to-next-to-leading logarithmic accuracy, so that order-$\as^2$
corrections are accounted for exactly, and logarithmically enhanced
contributions from loops of $b$ quarks are included.
This requires the calculation of the Mellin transform of the
order-$\as^2$ result in the whole complex plane for the Mellin
variable, which we provide for the first time for all the fragmenting
partons.  Evolution is performed to next-to-next-to-leading log
accuracy, and mixing with the gluon fragmentation function is taken
into account. The perturbative fragmentation functions are made available via LHAPDF grids.}

\arxivnumber{2207.10038}
\subheader{\small TIF-UNIMI-2022-11}

\begin{document}

\maketitle
\flushbottom

\input{sec-intro.tex}

\input{sec-largex2.tex}

\input{sec-results.tex}
\input{sec-conclusions.tex}

\section*{Acknowledgements}
We are grateful to Matteo Cacciari, Stefano Catani, Giancarlo Ferrera, Einan Gardi,
Alex Mitov, Ren\'e Poncelet, Gennaro Corcella and Paolo Nason for
useful discussions.  We thank Leonardo Bonino, Matteo Cacciari and Giovanni Stagnitto for having tested our code.
F.M.  is supported by the F.R.S.-FNRS through the IISN and the
``Excellence of Science'' EOS be.h project no. 30820817. 
The work of G.R. is supported by the Italian  Ministero dell’Università e della Ricerca (MUR) under grant PRIN 20172LNEEZ.
M.U. is supported by the European Research Council under the European Union’s Horizon 2020
research and innovation Programme (grant agreement n.950246). M.U. is
also partially supported by the Royal Society grant RGF/EA/180148 by the Royal Society grant
DH150088 and by the STFC consolidated grant ST/L000385/1. 
M.Z. is supported by the ``Programma per Giovani Ricercatori Rita Levi Montalcini'' granted by the Italian Ministero dell’Università e della Ricerca (MUR).
\newpage

\appendix
\input{app-res.tex}
\input{app-finite.tex}

\bibliography{paper.bbl}

\end{document}

%% file: sec-intro.tex

\section{Introduction}
The production of heavy quarks (charm, bottom, top), possibly in association with other particles,  is a particularly important class of processes at colliders. Not only they
provide key information for advancing our understanding of strong 
and electroweak interactions in the Standard Model (SM), but their very distinctive 
experimental signatures typically enter as background in many SM measurements and beyond-the-SM searches.   From the theoretical point of view, reliable and accurate predictions for these processes that match the current and foreseen experimental precision, are necessary. This is, however, a tall order. 
Fixed-order predictions are, in general, not accurate enough and terms appear that hamper the convergence of the perturbative series and need to be included at all orders.  A class of such terms involve mass logarithms, i.e. powers of $\log\frac{Q^2}{m^2}$, where $m$ and $Q$ are the heavy-quark mass and the typical energy scale of the process, respectively. When these terms are large, an all-order 
resummation, which can be achieved by means of Dokshitzer-Gribov-Lipatov-Altarelli-Parisi (DGLAP) evolution of fragmentation functions (FFs), becomes necessary. At past colliders and at the LHC this is important 
for the bottom and the charm quark. For top quarks such effects become relevant  at transverse momentum above a few TeV, a regime which is marginal at the LHC but possibly of interest at future high-energy colliders. FFs make it possible to organise terms of order $\as^p \log^q\frac{Q^2}{m^2}$ and to systematically resum them
via the DGLAP evolution, thus obtaining physical predictions at a given logarithmic accuracy. Nowadays,
the DGLAP evolution is implemented in several computer codes, such as
 {\tt QCDnum}~\cite{Botje:2010ay}, 
{\tt ffevol}~\cite{Hirai:2011si},  
{\tt APFEL}~\cite{Bertone:2013vaa}, 
 {\tt MELA}~\cite{Bertone:2015cwa}, 
 {\tt EKO}~\cite{Candido:2022tld},
 up to next-to-next-to-leading order (NNLO).

The fragmentation function of a $b$ quark is a special case: not only its dependence on the hard scale $Q$ is under control in perturbation theory via DGLAP  equations, but also the initial condition for evolution, typically given at a scale of the order of the $b$-quark mass, can be computed perturbatively~\cite{Mele:1990yq,Mele:1990cw,Cacciari:2001cw,Melnikov:2004bm,Mitov:2004du}. Such perturbative calculation becomes unreliable in the kinematical regime where the produced $b$ quark carries most of the available energy, and therefore the recoiling radiation is soft, giving rise to a further class of large logarithms. The formalism for the resummation of soft emission logarithms is outlined in~\cite{Cacciari:2001cw}, where resummation is explicitly performed to next-to-leading logarithmic (NLL) accuracy.

We stress that having an excellent perturbative description of FFs represents only part of the solution to the problem of providing an accurate description of heavy-quark 
related processes. The perturbative result has to be complemented  with a \emph{non-perturbative} part which parametrises the transition from the heavy quark to the
corresponding flavoured hadron. This contribution cannot be computed in perturbation theory, and is usually extracted by a direct comparison to data. Once the perturbative part is known, the extraction of the non-perturbative contribution to the FF  is a well-defined and independent phenomenological task. Though relevant to provide physical predictions to compare with experiment, it is not addressed in this work. The interested reader can find applications to the case of heavy-quark production in leptonic collisions in~\cite{Mele:1990cw,Cacciari:2001cw,Cacciari:2005uk,Aglietti:2006yf}, 
of the bottom quark in top decays in~\cite{Corcella:2001hz,Cacciari:2002re,Czakon:2021ohs}, and of the Higgs decay to heavy quarks in~\cite{Corcella:2004xv}.

In this work, we improve the perturbative description of the heavy-quark FF, using the framework to compute the coupled evolution of the $b$-quark fragmentation with the gluon and the other partons at NNLO accuracy described in~\cite{Ridolfi:2019bch} which builds upon {\tt MELA}~\cite{Bertone:2015cwa}. 
We asses  the impact of resummation of soft logarithms on the initial condition for the evolution of the $b$ fragmentation 
function $D_b(x,\mu^2_{0F})$ by implementing the resummation formalism of~\cite{Cacciari:2001cw} up to next-to-next-to-leading logarithmic accuracy (NNLL in the following).
Resummation at NNLL of the initial condition was already performed in~\cite{Aglietti:2006yf}. 
In that work the resummed result was matched to the exact perturbative calculation  at next-to-leading order (NLO).
We improve on this result in three ways. First, we compute the process-dependent component of the resummed initial condition in such a way to reproduce the exact order-$\as^2$ result of~\cite{Melnikov:2004bm} to NNLL accuracy, including the contributions originating from loops of heavy quarks. Note that such contributions had not been accounted for in previous works~\cite{Gardi:2005yi,Aglietti:2006yf}. Second, we match the NNLL resummed result to the exact NNLO initial condition provided in~\cite{Melnikov:2004bm}. This requires the calculation of the Mellin transform of the full NNLO initial condition, which was not available until now in an analitically-continued
version defined in the whole complex plane. The calculations presented here fill this gap. Finally, we include both the heavy-quark and the gluon and light quark FFs in their coupled evolution.

The work is organised as follows. 
In Sect.~\ref{sec:db} we summarise the formalism and the
relevant equations for the heavy-quark fragmentation function computation starting
from a perturbative initial condition. In particular, we focus on the dependence of the
results on the scheme used to deal with heavy flavours, and
 we present our calculation of the initial bottom fragmentation function
with soft logarithms resummed at NNLL and matched with the NNLO expression.
In Sect.~\ref{sec:results} we present selected numerical results, commenting on the impact of resummation as well as of the inclusion
of the NNLO initial conditions.
We present our conclusions in Sect.~\ref{sec:concl}.  Appendix~\ref{app:res} contains some details about the calculation of the resummed initial condition, while in Appendix~\ref{app:mellin} we illustrate the calculation of the relevant Mellin transforms.

%% file: sec-largex2.tex
\section{The heavy quark fragmentation function}
\label{sec:db}

In the fragmentation-function formalism, the cross section for the production of a heavy quark ${\mathcal Q}$ is given by
\beq
\sigma_{\mathcal Q}(x,Q^2,m^2)=\sum_j\int_x^1\frac{dz}{z}\,\hat\sigma_j\left(\frac{x}{z},\as(\mu_R^2),\mu_R^2,\mu_F^2,Q^2\right)D_{j{\mathcal Q}}(z,\mu_F^2,m^2)
+{\mathcal O}\left((m/Q)^p\right),
\eeq
where $x$ is the fraction of the available energy carried away by the produced heavy quark, $\hat \sigma_j$ is a hard cross section (possibly including parton distribution functions for the initial state) and $D_{j{\mathcal Q}}$ the fragmentation functions of partons $j$ into the heavy quark $\mathcal Q$.

The fragmentation functions at the scale $\mu_F$ are obtained by evolving suitable initial conditions, given at a reference scale $\mu_{0F}$, through Altarelli-Parisi equations. In the following, we will be interested in the initial condition in the case $j={\mathcal Q}=b$, which can be computed perturbatively at $\mu_{0F}\sim m$. We will denote by $D_b$ the $b$ fragmentation function for simplicity. We will also omit the explicit dependence of the fragmentation function on the $b$-quark mass $m$.

\subsection{Fixed-order calculation of the initial condition}

The initial condition for the heavy quark fragmentation function can
be computed perturbatively, as illustrated in~\cite{Mele:1990cw,Cacciari:2001cw}, at a
scale $\mu_{0F}$ of the order of (but not necessarily equal to~\cite{Bertone:2017djs}) the heavy quark mass. The order-$\as$ correction is given in~\cite{Mele:1990cw}, while the order-$\as^2$ coefficient was computed in~\cite{Melnikov:2004bm}. The result is
\beq
D_b^{\rm pert}(z,\mu_{0F}^2)=\delta(1-z)+\frac{\as^{(n_f)}(\mu_{0F}^2)}{2\pi}d_b^{(1)}(z,\mu_{0F}^2)
+\left(\frac{\as^{(n_f)}(\mu_{0F}^2)}{2\pi}\right)^2d_b^{(2)}(z,\mu_{0F}^2)+{\mathcal O}(\as^3),
\label{Dini}
\eeq
where
\begin{align}
d_b^{(1)}(z,\mu_{0F}^2)
&= C_F\left[\frac{1+z^2}{1-z}\left(\log\frac{\mu_{0F}^2}{m^2(1-z)^2}-1\right)\right]_+
\\
d_b^{(2)}(z,\mu_{0F}^2)
&= C_F^2 F^{(C_F^2)}_ {Q}(z,\mu_{0F}^2)
+C_AC_F F^{(C_AC_F)}_ {Q}(z,\mu_{0F}^2)
\nonumber\\
&+ C_FT_RF^{(C_FT_R)}_ {Q} (z,\mu_{0F}^2)+
C_FT_Rn_\ell F^{(C_FT_Rn_\ell)}_ {Q}(z,\mu_{0F}^2),
\label{d2Qresult}
\end{align}
and $n_\ell=n_f-1$ is the number of massless flavors. The functions  $F^{(C_F^2)}_ {Q}$,
$F^{(C_AC_F)}_ {Q}$, $F^{(C_FT_R)}_ {Q}$, $F^{(C_FT_Rn_\ell)}_ {Q}$ are given explicitly in~\cite{Melnikov:2004bm}. 
The calculation in~\cite{Melnikov:2004bm} is performed in the $\overline{\rm MS}$ renormalization scheme for ultraviolet divergences, which means that both massless and massive flavors take part in the evolution of the running coupling.

For the purpose of comparison with the resummed result, it will be useful to rewrite the result of~\cite{Melnikov:2004bm} as an expansion in powers of 
$\as=\as^{(n_\ell)}$. Furthermore, for greater generality it will be convenient to choose a renormalization scale $\mu_0$ different from $\mu_{0F}$. To order $\as^2$,
we have
\begin{align}
\as^{(n_f)}(\mu_{0F}^2)&=\as(\mu_0^2)-\as^2(\mu_0^2)
\left\{\left[b_0(n_\ell)-b_0(n_f)\right]\log\frac{m^2}{\mu_0^2}
+b_0(n_f)\log\frac{\mu_{0F}^2}{\mu_0^2}\right\}+\order{\as^3}
\nonumber\\
&=\as(\mu_0^2)-\as^2(\mu_0^2)
\left\{
\frac{T_f }{3\pi}
\log\frac{m^2}{\mu_0^2}
+\frac{11 C_A-4 T_f n_f}{12\pi}\log\frac{\mu_{0F}^2}{\mu_0^2}\right\}+\order{\as^3},
\label{alpha4}
\end{align}
where we have used
\beq
b_0(n_\ell)=\frac{11 C_A-4 T_f n_\ell}{12\pi};\qquad n_f=n_\ell+1.
\eeq
The first term in curly brackets originates from the change of renormalization scheme, while the second one arises from evolution from $\mu_0$ to $\mu_{0F}$.
Replacing Eq.~(\ref{alpha4}) in Eq.~(\ref{Dini})
and expanding again in powers of $\as(\mu_0^2)$ to order 2
we get
\begin{align}
&D_b^{\rm pert}(z,\mu_{0F}^2)=\delta(1-z)
+\frac{\as(\mu_0^2)}{2\pi}d_b^{(1)}(z,\mu_{0F}^2)
\nonumber\\
&+\left(\frac{\as(\mu_0^2)}{2\pi}\right)^2
\left\{d_b^{(2)}(z,\mu_{0F}^2)
-d_b^{(1)}(z,\mu_{0F}^2)\left[
\frac{2 T_f }{3}
\log\frac{m^2}{\mu_0^2}
+ \frac{11 C_A - 4 T_f (n_\ell+1)}{6}\log\frac{\mu_{0F}^2}{\mu_0^2}\right]
\right\}
\nonumber\\
&+{\mathcal O}(\as^3).
\end{align}
The calculation of the Mellin transform of this expression,
\beq
D_b^{\rm pert}(N,\mu_{0F}^2)=\int_0^1dz\,z^{N-1}D_b^{\rm pert}(z,\mu_{0F}^2)
\label{DpertN}
\eeq
as a function of the complex variable $N$ can be performed as illustrated in Appendix~\ref{app:mellin}; this is one of the main results of this paper.

\subsection{Soft gluon resummation}
It was observed in~\cite{Cacciari:2001cw} that logarithmic corrections to the initial condition for the heavy quark fragmentation function, arising at all orders because of soft emission, may play a relevant role, at least conceptually, in the large-$z$ region. It is therefore useful to resum such contributions to all orders, to some given logarithmic accuracy, and to assess their impact on the fragmentation function itself and on related observables. 
Soft gluon resummation of the $b$ fragmentation function at the initial scale was performed in~\cite{Cacciari:2001cw} to NLL accuracy, and subsequently improved to
NNLL in~\cite{Aglietti:2006yf}. Here we
  recompute the process-dependent component of the resummed initial
  condition, by including a term that was omitted in previous
  results, and improve on them by matching the resummed expression to the
order-$\as^2$ result of~\cite{Melnikov:2004bm}.

Soft gluon resummation is performed in Mellin space, as illustrated in Ref.~\cite{Cacciari:2001cw}. For a generic function $g(z)$, with
$0\le z\le 1$, we define
\beq
g(N)=\int_0^1dz\,z^{N-1}g(z).
\eeq
The resummed fragmentation function at the initial scale for evolution
$\mu_{0F}^2$ takes the familiar form of an exponential,
\beq
D_b^{\rm res}(N,\mu_{0F}^2)=g_0\left(\as(\mu_0^2),\frac{m^2}{\mu_0^2},\frac{\mu_{0F}^2}{\mu_0^2}\right)
\exp\left[ F\left(\as(\mu_0^2),\frac{m^2}{\mu_0^2},\frac{\mu_{0F}^2}{\mu_0^2},\lambda\right)\right],
\label{Dinires}
\eeq
where
\beq
F\left(\as(\mu_0^2),\frac{m^2}{\mu_0^2},\frac{\mu_{0F}^2}{\mu_0^2},\lambda\right)=\int_0^1dz\,\frac{z^{N-1}-1}{1-z}\left[\int^{\mu_{0F}^2}_{m^2(1-z)^2}\frac{d\mu^2}{\mu^2}\,A(\as(\mu^2))
+H(\as(m^2(1-z)^2))\right]
\label{expDinires}
\eeq
and
\beq
\lambda = b_0\as(\mu_0^2) \log N.
\eeq
The functions $A(\as)$ and $H(\as)$ have perturbative expansions in powers of 
$\as$ starting at order 1:
\beq
A(\as)=\sum_{k=1}^\infty A_k\left(\frac{\as}{\pi}\right)^k;\qquad H(\as)=\sum_{k=1}^\infty H_k\left(\frac{\as}{\pi}\right)^k.
\eeq
 NNLL accuracy is achieved including $A(\as)$ up to order $\as^3$ and $H(\as)$ up to order $\as^2$.
$A(\as)$ is determined by the Altarelli-Parisi splitting functions, while $H(\as)$ is process dependent, and must be extracted from the fixed-order calculation. The coefficients $H_i$ must be obtained by matching the expansion of the resummed expression Eq.~(\ref{Dinires}) to the relevant perturbative order with the fixed-order calculation.

In Eq.~(\ref{Dinires}) the strong coupling is computed with $n_\ell=4$ active flavour; this is the natural choice, because the integration over $\mu^2$ in the exponent
ranges between $m^2(1-z)^2\ll m^2$ and $\mu_{0F}^2\sim m^2$. Its evolution from $\mu_{0F}$ up to a hard scale $\mu_F\gg m$ is however performed with $n_f=n_\ell +1$ massless flavours.\footnote{We thank Stefano Catani for helping us clarifying this point.}

It is interesting to discuss the dependence of the resummed initial condition, Eq.~(\ref{Dinires}), on the initial factorization scale $\mu_{0F}$.
The fragmentation function evolved up to a generic hard scale $\mu_F$ is given by
\beq
D_b^{\rm res}(N,\mu_F^2)=E(N,\mu_F^2,\mu_{0F}^2)D_b^{\rm res}(N,\mu_{0F}^2),
\eeq
where 
\beq
E(N,\mu_F^2,\mu_{0F}^2)=E^{-1}(N,\mu_{0F}^2,\mu_F^2)
\eeq
is the Altarelli-Parisi evolution kernel at large $N$. We expect $D_b^{\rm res}(N,\mu_F^2)$ to be independent of
the initial scale $\mu_{0F}$, which implies
\begin{align}
\frac{\partial \log D_b^{\rm res}(N,\mu_{0F}^2)}{\partial\log\mu_{0F}^2}
&=-\frac{\partial}{\partial\log\mu_{0F}^2}\log E(N,\mu_F^2,\mu_{0F}^2)
\nonumber\\
&=\frac{\partial}{\partial\log\mu_{0F}^2}\log E(N,\mu_{0F}^2,\mu_F^2)
\nonumber\\
&=\gamma_{\rm AP}(N,\as^{(n_f)}(\mu_{0F}^2)),
\label{gamma01}
\end{align}
where 
\beq
\gamma_{\rm AP}(N,\as^{(n_f)}(\mu_{0F}^2))=-A(\as^{(n_f)}(\mu_{0F}^2))\log N+\order{N^0}
\eeq
 is the relevant $\overline{\rm MS}$ anomalous dimension in the large-$N$ limit. As already observed, the evolution between an initial scale $\mu_{0F}$, of the order or the heavy quark mass $m$, and a hard scale $\mu_F\gg m$, is determined by $n_f$ massless flavours;  the anomalous dimension is therefore naturally expressed as an expansion in powers of $\as^{(n_f)}$, whose
coefficients $A_k$ also depend on $n_f$ for $k\ge2$.
On the other hand, neglecting $\order{N^0}$ terms, the logarithmic derivative of Eq.~(\ref{Dinires}) reads
\begin{align}
&\frac{\partial \log D_b^{\rm res}(N,\mu_{0F}^2)}{\partial\log\mu_{0F}^2}
\nonumber\\
&=\int_0^1dz\,\frac{z^{N-1}-1}{1-z} \left[A(\as^{(n_\ell)}(\mu_{0F}^2))+\frac{\partial H(\as^{(n_\ell)}(m^2(1-z)^2))}{\partial\log\mu_{0F}^2}\right]+\order{N^0}
\nonumber\\
&=- \left[A(\as^{(n_\ell)}(\mu_{0F}^2))+\frac{\partial H(\as^{(n_\ell)}(m^2(1-z)^2))}{\partial\log\mu_{0F}^2}\right]\log N+\order{N^0}.
\label{gamma02}
\end{align}
Equations (\ref{gamma01}) and (\ref{gamma02}) can only be consistent with each other if the coefficients $H_k$ for $k\ge 2$ carry a dependence on $\log\frac{\mu_{0F}^2}{m^2}$.

After performing the two integrations in Eq.~(\ref{expDinires}), the exponent $F$ is expressed as a power expansion in $\as$ at fixed $\lambda$. To NNLL accuracy
\beq
F\left(\as(\mu_0^2),\frac{m^2}{\mu_0^2},\frac{\mu_{0F}^2}{\mu_0^2},\lambda\right)=\frac{1}{\as(\mu_0^2)}g_1(\lambda)+g_2(\lambda)+\as(\mu_0^2)g_3(\lambda)+{\rm N^3LL}.
\eeq
The functions $g_0,g_1,g_2,g_3$ are given explicitly in Appendix~\ref{app:res}, together with some details on their derivation, in terms of the coefficients
$A_1,A_2,A_3$ and $H_1,H_2$.
The functions $g_1(\lambda)$ and $g_2(\lambda)$ were already given in~\cite{Cacciari:2001cw}, and we find full agreement with those expressions. The function
$g_3(\lambda)$ was computed in~\cite{Aglietti:2006yf}, where NNLL resummation was performed, but not explicitly provided.

The extraction of $H_2$ from the order-$\as^2$ calculation deserves some comments. 
It can be obtained by expanding Eq.~(\ref{Dinires}) to order $\as^2$ and comparing the result with the fixed-order calculation in the large-$N$ limit, Eq.~(\ref{Dini}), after the replacement Eq.~(\ref{alpha4}). We obtain
\beq
H_2=-C_F \left[\frac{\pi b_0(n_\ell)}{9} + C_A\left(\frac{9 \zeta_3}{4} - \frac{\pi^2}{12} - \frac{11}{18}\right)\right]
+ \frac{C_F T_f}{54}\left[9\log^2\frac{m^2}{\mu_{0F}^2}+ 30\log\frac{m^2}{\mu_{0F}^2} +28\right],
\label{H2text}
\eeq
which depends on $\log\frac{\mu_{0F}^2}{m^2}$ as expected. 

We can check explicitly that, with $H_2$ given in Eq.~(\ref{H2}), the resummed initial condition $D_b^{\rm res}(N,\mu_{0F}^2)$ at NNLL is a solution of Eq.~(\ref{gamma01}). Indeed, neglecting constant (i.e.\ $N$ independent) terms,
\begin{align}
\frac{\partial \log D_b^{\rm res}(N,\mu_{0F}^2)}{\partial\log\mu_{0F}^2}
&=\frac{\partial}{\partial\log\mu_{0F}^2}F\left(\as(\mu_0^2),\frac{m^2}{\mu_0^2},\frac{\mu_{0F}^2}{\mu_0^2},\lambda\right)
\nonumber\\
&=-\left[\frac{\as(\mu_{0F}^2)}{\pi}A_1+\frac{\as^2(\mu_{0F}^2)}{\pi^2}A_2(n_\ell)+\frac{\as^2(\mu_{0F}^2)}{\pi^2}\frac{\partial H_2}{\partial\log\mu_{0F}^2}\right]\log N+\order{\as^3},
\end{align}
where we have taken into account that $H_1$ is $\mu_{0F}$-independent, and we have displayed the dependence of $A_2$ on the number of flavours $n_\ell$, see Eqs.~(\ref{A2},\ref{H2}). Since
\beq
\frac{\partial H_2}{\partial\log\mu_{0F}^2}=\frac{C_F T_f}{9}\left(3\log^2\frac{\mu_{0F}^2}{m^2}-5\right)
\eeq
we obtain
\begin{align}
&\frac{\partial \log D_b^{\rm res}(N,\mu_{0F}^2)}{\partial\log\mu_{0F}^2}
\nonumber\\
&=-\left\{\frac{A_1}{\pi}\left[\as(\mu_{0F}^2)+\as^2(\mu_{0F}^2)\frac{T_f}{3\pi}\log^2\frac{\mu_{0F}^2}{m^2}\right]
+\frac{\as^2(\mu_{0F}^2)}{\pi^2}\left[A_2(n_\ell)-\frac{5}{9}C_F T_f\right]
\right\}\log N+\order{\as^3},
\end{align}
where we have used $A_1=C_F$.
By Eq.~(\ref{alpha4}) with $\mu_0=\mu_{0F}$, 
\beq
\as(\mu_{0F}^2)+\as^2(\mu_{0F}^2)\frac{T_f}{3\pi}\log^2\frac{\mu_{0F}^2}{m^2}=\as^{(n_f)}(\mu_{0F}^2).
\eeq
Furthermore, from Eq.~(\ref{A2}),
\beq
A_2(n_\ell)-\frac{5}{9}C_F T_f=A_2(n_f).
\eeq
Hence
\begin{align}
\frac{\partial \log D_b^{\rm res}(N,\mu_{0F}^2)}{\partial\log\mu_{0F}^2}&=
-\left\{\frac{A_1}{\pi}\as^{(n_f)}(\mu_{0F}^2)
+\frac{\as^2(\mu_{0F}^2)}{\pi^2}A_2(n_f)
\right\}\log N+\order{\as^3}
\nonumber\\
&=\gamma_{\rm AP}(N,\as^{(n_f)}(\mu_{0F}^2))
\end{align}
as announced.

Our result, Eq.~(\ref{H2text}), differs from  the value of $H_2$ given in~\cite{Gardi:2005yi} by the last term, proportional to $C_F T_f$ and $\mu_{0F}$-dependent.
This extra term is different from zero for all choices of $\mu_{0F}$.
We have seen that a dependence of $H_2$ on the initial scale $\mu_{0F}$ is expected on general grounds, and can be read off the Altarelli-Parisi anomalous dimension
in the large-$N$ limit. We have also shown that our result Eq.~(\ref{H2text}) is consistent with expectations.
In~\cite{Gardi:2005yi}, $H_2$ was extracted under the assumption that the $b$ quark appears only as an external line, i.e.\ no loops involving $b$ quarks was considered;
here, we do not make this assumption.
The value of $H_2$ obtained in~\cite{Gardi:2005yi} was later employed in~\cite{Aglietti:2006yf}. 

\subsection{Matching resummed and fixed-order initial condition}

Both the fixed-order initial condition for the $b$ fragmentation function Eq.~(\ref{DpertN}), accurate to NNLO, and the resummed initial condition Eq.~(\ref{Dinires}), accurate to NNLL, are now expressed as functions of $\as(\mu_0^2)=\as^{(n_\ell)}(\mu_0^2)$, and of the complex variable $N$, Mellin conjugate to $z$. Their combination requires the subtraction of the resummed result expanded to order $\as^2$ to avoid double counting. Our final result is therefore
\begin{align}
D_b(N,\mu_{0F}^2)&=D_b^{\rm pert}(N,\mu_{0F}^2)+D_b^{\rm res}(N,\mu_{0F}^2)
\nonumber\\
&-\left[1+\as(\mu_0^2)\left.
\frac{\partial D_b^{\rm res}(N,\mu_{0F}^2)}{\partial\as}\right|_{\as=0}
+\frac{1}{2}\as^2(\mu_0^2)\left.
\frac{\partial^2 D_b^{\rm res}(N,\mu_{0F}^2)}{\partial\as^2}\right|_{\as=0}\right].
\end{align}

%% file: sec-results.tex
\section{Results}
\label{sec:results}
In this Section, we present results for the $b$-quark fragmentation function $D_b(z,\mu_F^2)$ and its Mellin transform $D_b(N,\mu_F^2)$. Our aim is to assess the impact of the
NNLL soft-gluon resummation for the initial condition, as well as the size of $\mathcal O (\as^2)$ terms in the latter.
We will always employ NNLL DGLAP evolution, and we will set the following values for the bottom mass and the factorisation scale:
\begin{equation}
m = 4.7\, \gev,\qquad \mu_F=100\,\gev.
\end{equation}
For all results presented in this section we will set $\mu_{0F}=\mu_0$.
Specifically, we will consider two values for the initial scale $\mu_0$, namely $\mu_0=m$ (displayed in the left panels)
and $\mu_0=2 m$ (in the right panels). 

The Mellin-transfomed initial condition $D_b(N,\mu_0^2)$ displays a logarithmic branch cut on the positive $N$ real axis for $\lambda>\frac{1}{2}$, or
\beq
N>N_L(\mu_0)=e^{\frac{1}{2 b_0\as(\mu_0^2)}}
\eeq
originated by the Landau singularity of the running coupling. The resummed initial condition is meaningless for $N$ too close to $N_L(\mu_0)$. We find
\beq
N_L(m)\sim 35;\qquad N_L(2m)\sim 65.
\eeq
For this reason, we choose a different range in $N$ for the two values of $\mu_0$.

\begin{figure}[ht!]
    \centering
    \includegraphics[width=\textwidth]{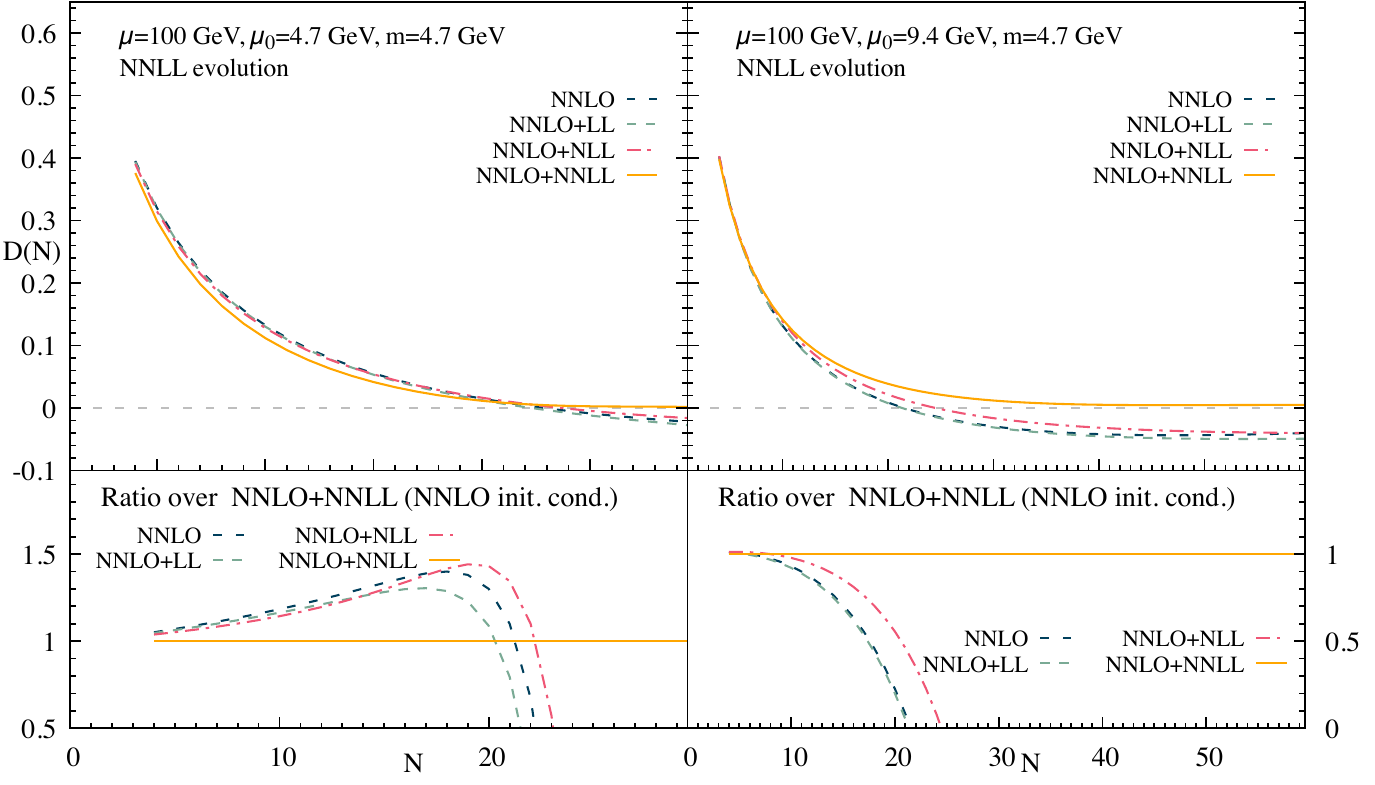}
    \caption{\label{fig:reslx-DbN} Predictions for the bottom-quark FF as a function of $N$ in the positive real axis, including soft resummation up to NNLL and initial conditions at NNLO.}
\end{figure}
\begin{figure}[ht!]
    \centering
    \includegraphics[width=\textwidth]{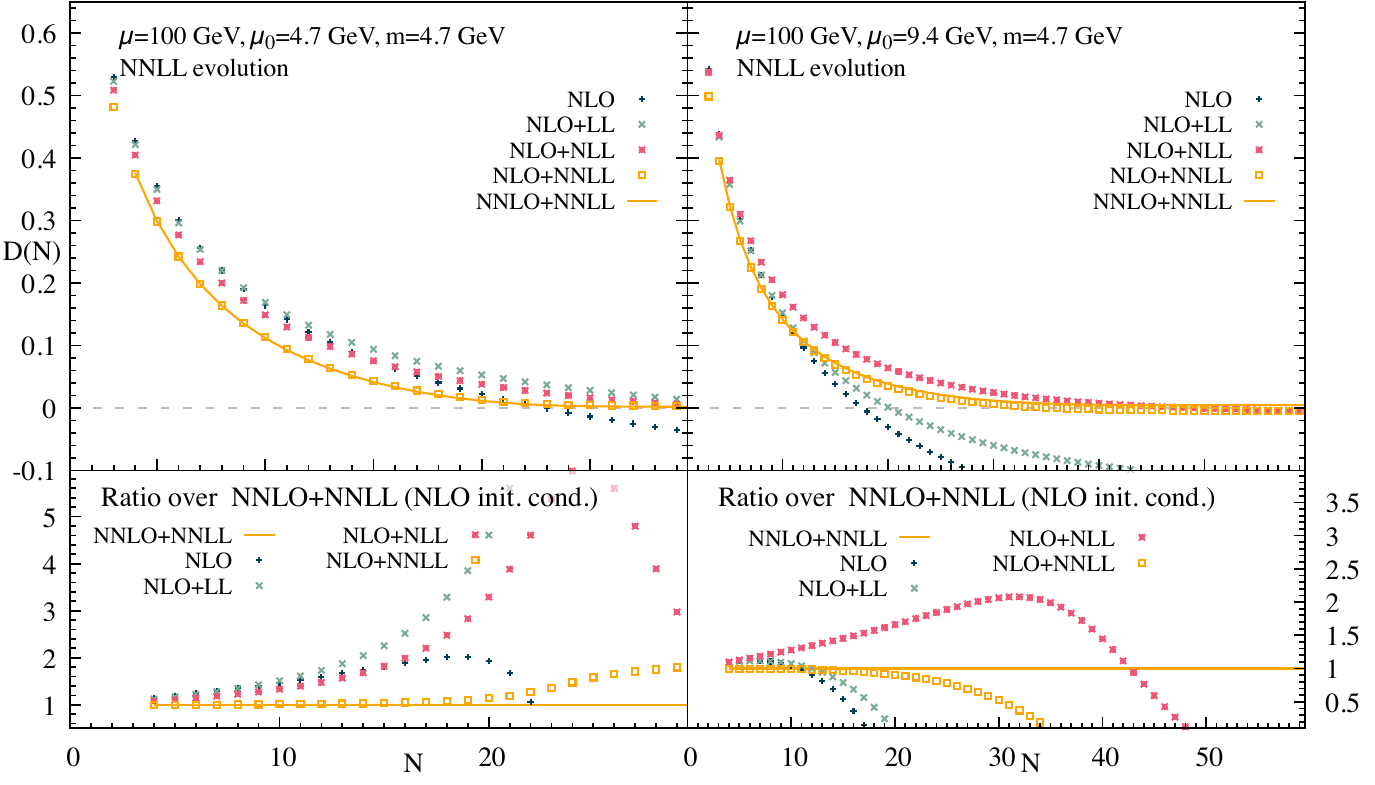}
    \caption{\label{fig:reslx-DbN-noas2} Predictions for the bottom-quark FF as a function of $N$ in the positive real axis, including soft resummation up to NNLL and initial conditions at NLO. 
The prediction indluding NNLL resummation and NNLO initial conditions is also shown as reference.}
\end{figure}

In Fig.~\ref{fig:reslx-DbN} we show results for $D_b(N,\mu_F^2)$.
We include soft resummation at different logarithmic accuracies on top of the NNLO initial conditions: the dashed black curve shows
the predictions without resummation, i.e.\ the same result presented in Ref.~\cite{Ridolfi:2019bch}, which is seen to become negative at $N\sim 20$. LL, NLL and NNLL resummation are included in the dashed teal, dot-dashed magenta and solid orange curves, respectively. It can be appreciated that the NNLL resummed prediction is the only one which remains positive over
the whole considered $N$ range, while all other predictions become negative for sufficiently large values of $N$. Also, at large $N$ ($N>10$), effects of resummation remain 
quite important at all considered orders. Finally, when the initial scale $\mu_0$ is increased by a factor 2, differences among the predictions turn larger, a fact related
to the smaller impact of the DLGAP evolution, which is common to all predictions, in favour of that of the initial conditions.

The impact of the NNLO initial condition can be better appreciated in Fig.~\ref{fig:reslx-DbN-noas2}, where we show the same predictions as in Fig.~\ref{fig:reslx-DbN},
 but obtained including only NLO initial conditions. The different predictions are displayed as symbols, with the same color code as in Fig.~\ref{fig:reslx-DbN} for the corresponding logarithmic accuracy, 
 and compared to the NNLO+NNLL one. We observe that i) resummation effects play an even more important role in this case; otherwise stated, the inclusion of NNLO initial condition has a stabilising effect in this respect; ii) with NLO initial conditions, NLL resummation is sufficient to get positive-definite predictions up to $N\approx 50$; iii) if NNLL resummation is included,
 the effect of finite terms in the NNLO initial conditions is small for $N<20$,
 while it amounts to several 10\%s for larger values of $N$.
 \begin{figure}[ht!]
    \centering
    \includegraphics[width=\textwidth]{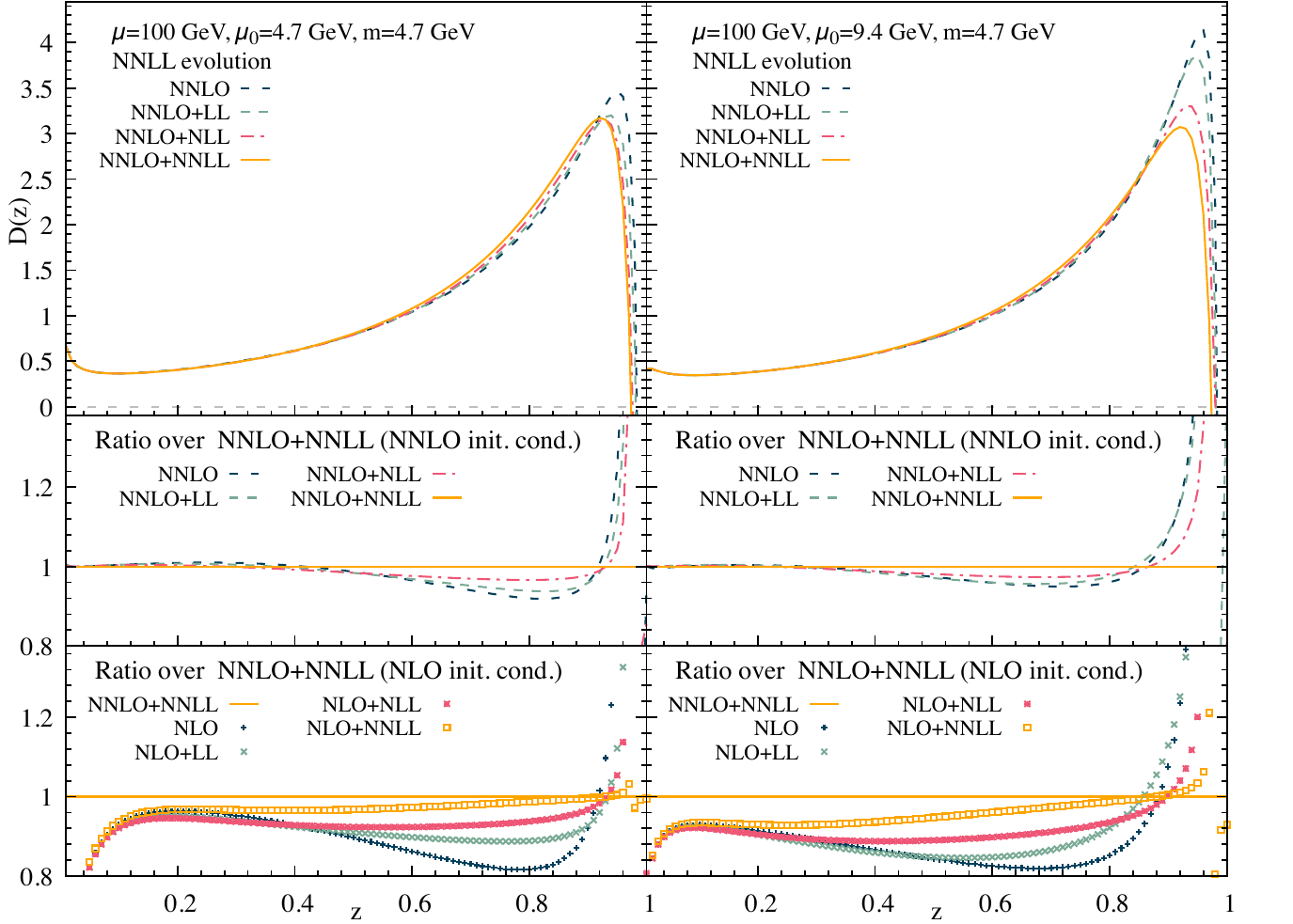}
    \caption{\label{fig:reslx-Dbx}Predictions for the bottom-quark FF in $z$ space, including soft resummation up to NNLL and initial conditions at NNLO or NLO.}
\end{figure}

Turning to predictions in the physical space of momentum fraction $z$, Fig.~\ref{fig:reslx-Dbx} provides a global view of the effects, where the same patterns as in Figs.~\ref{fig:reslx-DbN} and \ref{fig:reslx-DbN-noas2} are employed for the corresponding predictions.
 The most visible effect of resummation is to reduce the value of the FF at the large-$z$ peak, slightly moving its position $z_P$
 to the left, and slightly raising the tail for $z<z_P$. Unlike in $N$ space, predictions generally reflect the hierarchy of the resummation orders, both when NNLO and NLO initial conditions are employed (central and lower inset
 respectively). The hierarchy is violated  only for very large values of $z$ (well right of $z_P$), where the differences at large $N$ are reflected.
 Considering the effects of including NNLO initial conditions, by comparing the central and lower inset we see that their inclusion reduces the impact of resummation, as already observed in $N$ space. We also notice
 that, when NNLL resummation is employed, genuine $\mathcal O(\as^2)$ effects are generally at the percent level (although not uniformly in the $z$ range), 
 about 5\% for $\mu_0=m$ and $z<z_P$, and twice as much when $\mu_0$ is increased to $2m$.

%% file: sec-conclusions.tex
\section{Conclusions}
\label{sec:concl}
In this work, we have obtained results for the $b$-quark fragmentation function including NNLL soft-gluon resummation for the perturbative initial conditions matched with the NNLO result.

Following the approach of~\cite{Cacciari:2001cw}, we have performed the resummation of soft emission logarithms for the Mellin-transformed initial condition. 
First, we have identified a term in the process-dependent component of the NNLL resummed initial condition that was omitted in previous works. Note that the inclusion of this term makes the resummed result fully consistent with the exact NNLO perturbative calculation and with the DGLAP evolution. 
Moreover we have computed the Mellin transform of the full NNLO initial condition, and obtained its analytic continuation to the complex plane of the variable $N$, Mellin-conjugate to the momentum fraction $z$, which was not available in the existing literature. Finally we have considered the coupled evolution of the quark and gluon FFs.

We find that NNLL resummation improves the predictions obtained by evolving initial conditions at NNLO. Their behaviour  is regular across the relevant range of the Mellin variable $N$, similarly to what happens in the NLO+NLL case. The inclusion of NNLO initial conditions provides a sizeable stabilisation of resummed predictions in $z$ space, and has a moderate, yet appreciable effect on the
evolved fragmentation function. Our work makes it possible to include the NNLO+NNLL initial conditions, convolved with a suitable short-distance cross section, in a wide range of phenomenological applications related to heavy-hadron production at colliders and in particular at the LHC. 
The perturbative fragmentation functions are available via LHAPDF grids and can be requested to the authors. An easy-to-use computer code implementing the NNLO+NNLL initial conditions supplemented by NNLO DGLAP evolution is in preparation.

%% file: app-res.tex
\section{Calculation of the resummed initial condition for $D_b$}
\label{app:res} 
In this Appendix we describe in detail the calculation of
the resummed $b$ fragmentation function at the initial scale $\mu_{0F}$, Eq.~(\ref{Dinires}):
\beq
D_b^{\rm res}(N,\mu_{0F}^2)=g_0\left(\as(\mu_0^2),\frac{m^2}{\mu_0^2},\frac{\mu_{0F}^2}{\mu_0^2}\right)
\exp\left[ F\left(\as(\mu_0^2),\frac{m^2}{\mu_0^2},\frac{\mu_{0F}^2}{\mu_0^2},\lambda\right)\right],
\eeq
where
\beq
 \lambda=b_0\as(\mu_0^2) \log N,
\eeq
to NNLL accuracy. In this expression, $\as=\as^{(n_\ell)}$ is defined in the $n_\ell=n_f-1$ renormalization scheme:
\beq
\mu^2\frac{d\as(\mu^2)}{d\mu^2}=\beta(\as)=-b_0\as^2-b_1 \as^3-b_2 \as^4+{\mathcal O}(\as^5),
\eeq
where
\begin{align}
b_0&=b_0(n_\ell)= \frac{11 C_A - 4 T_f n_\ell}{12\pi},
\label{b0}
\\
b_1&=b_1(n_\ell) = \frac{17 C_A^2 - 10 C_A T_f n_\ell - 6 C_F T_f n_\ell}{24\pi^2},
\\
b_2&=b_2(n_\ell)= \frac{1}{128\pi^3}\left(2857 - \frac{5033}{9} n_\ell+ \frac{325}{27}n_\ell^2\right).
\end{align}
The renormalization scale $\mu_0$ is taken to be
different in general from the initial factorization scale $\mu_{0F}$. Both are taken of the order of the heavy quark mass $m$.

We first consider the exponent $F$. We have
\begin{align}
&F\left(\as(\mu_0^2),\frac{m^2}{\mu_0^2},\frac{\mu_{0F}^2}{\mu_0^2},\lambda\right)=\int_0^1dz\,\frac{z^{N-1}-1}{1-z}f\left(\as(\mu_0^2),\frac{m^2}{\mu_0^2},\frac{\mu_{0F}^2}{\mu_0^2},\lambda_z\right)
\\
&f\left(\as(\mu_0^2),\frac{m^2}{\mu_0^2},\frac{\mu_{0F}^2}{\mu_0^2},\lambda_z\right)=
\int^{\mu_{0F}^2}_{m^2(1-z)^2}\frac{d\mu^2}{\mu^2}\,A(\as(\mu^2))+H(\as(m^2(1-z)^2)),
\end{align}
where
\beq
 \lambda_z=-b_0\as(\mu_0^2) \log(1-z).
\eeq
To NNLL accuracy, 
\begin{align}
&A(\as)=A_1\frac{\as^{\rm NNLL}}{\pi}
+A_2\left(\frac{\as^{\rm NLL}}{\pi}\right)^2
+A_3\left(\frac{\as^{\rm LL}}{\pi}\right)^3,
\end{align}
where
\begin{align}
A_1 &= C_F
\label{A1}
\\
A_2 &= C_F \left[\frac{5}{3} \pi b_0 + C_A \left(\frac{1}{3} - \frac{\pi^2}{12}\right)\right]
\label{A2}
\\
A_3 &= C_F \Bigg\{-\frac{\pi^2 b_0^2}{3} + 
     \pi b_0 \left[\left(\frac{55}{16} - 3 \zeta_3\right) C_F 
     + \left(\frac{253}{72} - \frac{5 \pi^2}{18} + \frac{7\zeta_3}{2}\right) C_A\right]
           \nonumber\\
     &+ \left[
     \left(-\frac{605}{192} + \frac{11\zeta_3}{4}\right) C_A C_F + \left(-\frac{7}{18} - \frac{\pi^2}{18} - \frac{11\zeta_3}{4} + 
          \frac{11 \pi^4}{720}\right) C_A^2\right]\Bigg\},
\end{align}
and
\beq
H(\as)=H_1\frac{\as^{\rm NLL}}{\pi}+H_2\left(\frac{\as^{\rm LL}}{\pi}\right)^2,
\label{Hterm}
\eeq
where
\begin{align}
H_1 &= -C_F
\label{H2}
\\
H_2&=-C_F \left[\frac{\pi b_0}{9} + C_A\left(\frac{9 \zeta_3}{4} - \frac{\pi^2}{12} - \frac{11}{18}\right)\right]
+ \frac{C_F T_f}{54}\left[9\log^2\frac{m^2}{\mu_{0F}^2}+ 30\log\frac{m^2}{\mu_{0F}^2} +28\right].
\end{align}
The $\mu^2$ integration of the $A$ term is conveniently performed in the variable
$\alpha=\as(\mu^2)$:
\beq
\frac{d\mu^2}{\mu^2}=\frac{d\alpha}{\beta(\alpha)}.
\eeq
To NNLL accuracy,
\begin{align}
f\left(\as(\mu_0^2),\frac{m^2}{\mu_0^2},\frac{\mu_{0F}^2}{\mu_0^2},\lambda_z\right)
&=\frac{A_1}{\pi}\int_{\alpha^{\rm NNLL}_-}^{\alpha^{\rm NNLL}_+}\frac{d\alpha}{b_0\alpha+b_1\alpha^2+b_2\alpha^3}
+\frac{A_2}{\pi^2}\int_{\alpha^{\rm NLL}_-}^{\alpha^{\rm NLL}_+}\frac{d\alpha}{b_0+b_1\alpha}
+\frac{A_3}{\pi^3}\int_{\alpha^{\rm LL}_-}^{\alpha^{\rm LL}_+}\frac{\alpha^2d\alpha}{b_0}
\nonumber\\
&+\frac{H_1}{\pi}\as^{\rm NLL}(m^2(1-z)^2)+\frac{H_2}{\pi^2}\left(\as^{\rm LL}(m^2(1-z)^2)\right)^2
\label{Aterm}
\end{align}
where
\beq
\alpha^{\rm N^kLL}_-=\as(\mu_{0F}^2);\qquad\alpha^{\rm N^kLL}_+=\as(m^2(1-z)^2),
\label{apm}
\eeq
and
\begin{align}
\as^{\rm LL}(\mu^2)&=\frac{\as(\mu_0^2)}{X(\mu^2)}
\nonumber\\
\as^{\rm NLL}(\mu^2)&=\as^{\rm LL}(\mu^2) - \frac{b_1}{b_0}\left[ \frac{\as(\mu_0^2)}{X(\mu^2)} \right]^2
\log X(\mu^2)
\nonumber\\
\as^{\rm NNLL}(\mu^2)&=\as^{\rm NLL}(\mu^2)
+\left[ \frac{\as(\mu_0^2)}{X(\mu^2)} \right]^3\Bigg\{
\left(\frac{b_1}{b_0}\right)^2
\left[\log^2 X(\mu^2) - \log X(\mu^2) - 1 + X(\mu^2)\right] 
\nonumber\\
&+ \frac{b_2}{b_0} (1 - X(\mu^2));
\Bigg\}
\\
X(\mu^2)&=1+b_0 \as(\mu_0^2)\log\frac{\mu^2}{\mu_0^2}.
\end{align}
The integrals in eq.~(\ref{Aterm}) are easily computed, and the result is an analytic function of $\lambda_z$:
\beq
f\left(\as(\mu_0^2),\frac{m^2}{\mu_0^2},\frac{\mu_{0F}^2}{\mu_0^2},\lambda_z\right)=\sum_{p=0}^\infty f_p\lambda_z^p=\sum_{p=0}^\infty f_p(-b_0\as(\mu_0^2))^p \log^p(1-z).
\eeq
Hence
\beq
F\left(\as(\mu_0^2),\frac{m^2}{\mu_0^2},\frac{\mu_{0F}^2}{\mu_0^2},\lambda\right)=\sum_{p=0}^\infty f_p(-b_0\as(\mu_0^2))^p I_p,
\eeq
where
\beq
I_p=\int_0^1 dz\,\frac{z^{N-1}-1}{1-z}\,\log^p(1-z).
\eeq
The functions $I_p$ can be computed by taking derivatives of a generating function:
\beq
I_p=\lim_{\eta\to 0}\frac{d^p}{d\eta^p}\,G(\eta),
\label{ipgen}
\eeq
where
\beq
G(\eta)=\int_0^1 dz\,(z^{N-1}-1)\,(1-z)^{\eta-1}
=\frac{\Gamma(N)\Gamma(\eta)}{\Gamma(N+\eta)}-\frac{1}{\eta}
=\frac{1}{\eta}\left[\Gamma(1+\eta)N^{-\eta}-1\right]
+{\mathcal O}(1/N).
\eeq
In the last step we have used the Stirling approximation for the $\Gamma$ function at large values of its argument, which is the limit we are interested in.
We obtain 
\beq
\label{firstres}
I_p=\frac{1}{p+1}\sum_{k=0}^{p+1}\binom{p+1}{k}\,\Gamma^{(k)}(1)\,L^{p+1-k}
+{\mathcal O}(1/N),
\eeq
where
\beq
L=\log\frac{1}{N}.
\eeq
We now observe that
\beq
\label{dll}
\frac{1}{p+1}\,k!\,\binom{p+1}{k}\,L^{p+1-k}=-\frac{d^k}{dL^k}\int_0^{1-\frac{1}{N}}dz\,\frac{\log^p(1-z)}{1-z}.
\eeq
Replacing eq.~(\ref{dll}) in (\ref{firstres}) we obtain 
\beq
\label{secondres}
I_p=-\sum_{k=0}^\infty
\frac{\Gamma^{(k)}(1)}{k!}
\frac{d^k}{dL^k}\int_0^{1-\frac{1}{N}}dz\,\frac{\log^p(1-z)}{1-z}
+{\mathcal O}(1/N).
\eeq
Note that the sum has been extended to $\infty$ because all derivatives of order $p+2$ and higher vanish.
We have therefore
\begin{align}
F\left(\as(\mu_0^2),\frac{m^2}{\mu_0^2},\frac{\mu_{0F}^2}{\mu_0^2},\lambda\right)&=\sum_{p=0}^\infty f_p(-b_0\as(\mu_0^2))^p I_p
\nonumber\\
&=-\sum_{p=0}^\infty f_p(-b_0\as(\mu_0)^2)^p
\sum_{k=0}^\infty
\frac{\Gamma^{(k)}(1)}{k!}
\frac{d^k}{dL^k}\int_0^{1-\frac{1}{N}}dz\,\frac{\log^p(1-z)}{1-z}
\nonumber\\
&=-
\sum_{k=0}^\infty
\frac{\Gamma^{(k)}(1)}{k!}
\frac{d^k}{dL^k}\int_0^{1-\frac{1}{N}}dz\,\frac{f(\lambda_z)}{1-z}.
\end{align}
The $z$ integration is conveniently performed in the variable $\lambda_z$:
\beq
d\lambda_z=b_0\as(\mu_0^2)\frac{dz}{1-z};
\qquad
0<\lambda_z<\lambda.
\eeq
Furthermore
\beq
\frac{d^k}{dL^k}=(-b_0\as(\mu_0^2))^k\frac{d^k}{d\lambda^k}.
\eeq
We find
\beq
F\left(\as(\mu_0^2),\frac{m^2}{\mu_0^2},\frac{\mu_{0F}^2}{\mu_0^2},\lambda\right)=
-\frac{1}{b_0\as(\mu_0^2)}\sum_{k=0}^\infty
\frac{\Gamma^{(k)}(1)}{k!}(-b_0\as(\mu_0^2))^k
\frac{d^k}{d\lambda^k}\int_0^\lambda d\lambda_z\,f(\lambda_z).
\eeq
An even simpler expression is obtained by separating off the first term in the sum, which is the only one which requires an integration:
\begin{align}
F\left(\as(\mu_0^2),\frac{m^2}{\mu_0^2},\frac{\mu_{0F}^2}{\mu_0^2},\lambda\right)&=
-\frac{1}{b_0\as(\mu_0^2)}
\int_0^\lambda d\lambda_z\,f(\lambda_z)
+\sum_{k=1}^\infty
\frac{\Gamma^{(k)}(1)}{k!}(-b_0\as(\mu_0^2))^{k-1}
\frac{d^k}{d\lambda^k}\int_0^\lambda d\lambda_z\,f(\lambda_z)
\nonumber\\
&=
-\frac{1}{b_0\as(\mu_0^2)}
\int_0^\lambda d\lambda_z\,f(\lambda_z)
+\sum_{k=0}^\infty
\frac{\Gamma^{(k+1)}(1)}{(k+1)!}(-b_0\as(\mu_0^2))^k
\frac{d^k}{d\lambda^k}f(\lambda).
\end{align}
To NNLL accuracy, only the terms $k=0,1$ are relevant in the sum. We therefore obtain our final expression
\beq
F\left(\as(\mu_0^2),\frac{m^2}{\mu_0^2},\frac{\mu_{0F}^2}{\mu_0^2},\lambda\right)=
-\frac{1}{b_0\as(\mu_0^2)}
\int_0^\lambda d\lambda_z\,f(\lambda_z)
+\Gamma^{(1)}(1)f(\lambda)
-\frac{\Gamma^{(2)}(1)}{2}b_0\as(\mu_0^2)\frac{d}{d\lambda}f(\lambda)+\rm N^3LL
\eeq
which can be cast in the form of an expansion in powers of $\as(\mu_0^2)$ at fixed $\lambda$:
\beq
F\left(\as(\mu_0^2),\frac{m^2}{\mu_0^2},\frac{\mu_{0F}^2}{\mu_0^2},\lambda\right)=\frac{1}{\as(\mu_0^2)}g_1(\lambda)+g_2(\lambda)+\as(\mu_0^2)g_3(\lambda)+\rm N^3LL
\eeq
(we have omitted the dependence of $g_2,g_3$ on $\frac{m^2}{\mu_0^2},\frac{\mu_{0F}^2}{\mu_0^2}$ for simplicity.)

Constant terms, i.e. $N$-independent terms, are not controlled by Sudakov resummation. Note that, because the expansion of $F$ starts at order $\as$, we have
\begin{align}
g_1(\lambda)&=g_1^{(2)}\lambda^2+g_1^{(3)}\lambda^3+{\mathcal O}(\lambda^4)
\\
g_2(\lambda)&=g_2^{(1)}\lambda+g_1^{(2)}\lambda^2+{\mathcal O}(\lambda^3)
\\
g_3(\lambda)&=g_3^{(0)}+g_1^{(1)}\lambda+{\mathcal O}(\lambda^2).
\end{align}
So, $N$-independent terms in $F$ first appear at NNLL. We remove them by replacing
\beq
g_3(\lambda)\to g_3(\lambda)-g_3^{(0)}.
\eeq
We find
\begin{align}
g_1(\lambda)&=-\frac{A_1}{2\pi b_0^2}\left[2 \lambda + (1 - 2 \lambda) \log(1 - 2\lambda)\right]
\\
g_2(\lambda)&=\frac{1}{4\pi^2 b_0^3}\Bigg\{  
4\lambda\left[A_2 b_0 - A_1 \left(b_1 + b_0^2\log\frac{\mu_{0F}^2}{\mu_0^2}\right) \pi\right] 
\nonumber\\
&+ 
 2 \left[A_2 b_0 + b_0^2 H_1 \pi - 
    A_1 \left(b_1 + b_0^2 \left(\log\frac{m^2}{\mu_0^2}-2\gamma \right)\right) \pi\right] \log(1 - 2\lambda) 
    \nonumber\\
    &-  A_1 b_1 \pi \log^2 (1 - 2\lambda)\Bigg\},
\end{align}
in agreement with the results presented in ref.~\cite{Cacciari:2001cw}.
We also find
\begin{align}
g_3(\lambda)&=-\frac{1}{12\pi^3 b_0^4(1-2\lambda)}
\Bigg\{2\lambda\Big[6 A_3 b_0^2 \lambda 
\nonumber\\
&+ \pi\Big[6 b_0^3 H_2 + 6 A_2 \left[-b_0 b_1(1 + \lambda)+ b_0^3\left(2\gamma - \log\frac{m^2}{\mu_0^2}+ (1 - 2 \lambda)\log\frac{\mu_{0F}^2}{\mu_0^2} \right)\right]
\nonumber\\
&+6 A_1 b_0 b_2 (1 -\lambda) \pi + 6 A_1 b_1^2\lambda \pi - 6 b_0^2 b_1 \left[A_1\left(2 \gamma -\log\frac{m^2}{\mu_0^2}\right)+ H_1 \right] \pi 
\nonumber\\
&+ b_0^4 \pi 
\left[6 H_1 \left(2\gamma - \log\frac{m^2}{\mu_0^2}\right) 
+ A_1 \left(3\left(2\gamma-  \log\frac{m^2}{\mu_0^2} \right)^2 -3 (1 -2\lambda) \log^2\frac{\mu_{0F}^2}{\mu_0^2} + 2 \pi^2\right)
             \right]
             \Big]\Big]
\nonumber\\
&- 6 \pi \left[A_2 b_0 b_1 + b_0^2 b_1 H_1 \pi 
- A_1 \left(2 b_1^2\lambda + b_0 b_2 (1 - 2\lambda) 
    - b_0^2 b_1 \left(2\gamma - \log\frac{m^2}{\mu_0^2}\right)\right) \pi\right]\log(1 - 2\lambda) 
    \nonumber\\
    &+  3 A_1 b_1^2 \pi^2 \log^2(1 - 2\lambda)
\Bigg\}.
\end{align}
This is a new result: NNLL resummation was performed in~\cite{Aglietti:2006yf}, but an explicit expression of $g_3(\lambda)$ is not given there.

We now turn to the pre-exponential factor $g_0$ in eq.~(\ref{Dinires}).
The function $g_0$ is defined in such a way that constant terms are correctly taken into account up to order in $\as^2$. 
It can therefore be read off the result of Ref.~\cite{Melnikov:2004bm}:
\beq
g_0\left(\as(\mu_0^2),\frac{m^2}{\mu_0^2},\frac{\mu_{0F}^2}{\mu_0^2}\right)=1
+\frac{\as^{(n_f)}(\mu_{0F}^2)}{2\pi}d_b^{(1,c)}
+\left(\frac{\as^{(n_f)}(\mu_{0F}^2)}{2\pi}\right)^2d_b^{(2,c)},
\label{g0}
\eeq
where
\beq
\as^{(n_f)}(\mu_{0F}^2)=\as(\mu_0^2)-\as^2(\mu_0^2)
\left\{\left[b_0(n_\ell)-b_0(n_f)\right]\log\frac{m^2}{\mu_0^2}
+ b_0(n_f)\log\frac{\mu_{0F}^2}{\mu_0^2}\right\}+{\mathcal O}(\as^3)
\eeq
and
\beq
d_b^{(1,c)}=-C_F\left[2\gamma^2 + \frac{3}{2}\log\frac{m^2}{ \mu_{0F}^2} - 2\gamma \left(1 + \log\frac{m^2}{ \mu_{0F}^2}\right) -
   2 \left(1 - \frac{\pi^2}{6}\right)\right]
\eeq

\begin{align}
d_b^{(2,c)}&=C_A C_F \Bigg[\left(\frac{11}{8}-\frac{11 \gamma }{6}\right) \log^2\frac{m^2}{ \mu_{0F}^2}
-\left(-3 \zeta (3)+\frac{35}{8}-\frac{34 \gamma}{9}-\frac{11 \gamma ^2}{3}+\frac{\gamma  \pi ^2}{3}\right)\log\frac{m^2}{ \mu_{0F}^2} 
\nonumber\\
&\qquad+9 \gamma  \zeta (3)-\frac{97 \zeta (3)}{18}+\frac{\pi ^4}{12}+\frac{\gamma ^2 \pi
   ^2}{3}-\frac{14 \gamma  \pi ^2}{9}+\frac{7 \pi ^2}{54}-\frac{22 \gamma ^3}{9}-\frac{34 \gamma ^2}{9}-\frac{55 \gamma }{27}+\frac{1141}{288}+\pi
   ^2 \log 2\Bigg]
   \nonumber\\
   &+C_F^2 \Bigg[\frac{1}{8} (4 \gamma -3)^2 \log^2\frac{m^2}{ \mu_{0F}^2}
   -\left(6 \zeta (3)+\frac{27}{8}-\gamma -7 \gamma ^2+4\gamma ^3-\pi ^2+\frac{2 \gamma  \pi ^2}{3}\right)\log\frac{m^2}{ \mu_{0F}^2} 
   \nonumber\\
   &\qquad-\frac{3 \zeta (3)}{2}-\frac{11 \pi ^4}{180}+\frac{2 \gamma ^2 \pi ^2}{3}-\frac{2 \gamma  \pi^2}{3}+\frac{\pi ^2}{4}+2 \gamma ^4-4 \gamma ^3-2 \gamma ^2+4 \gamma +\frac{241}{32}-2 \pi ^2 \log 2\Bigg]
   \nonumber\\
   &+C_F T_R
   \Bigg[\left(\frac{2 \gamma }{3}-\frac{1}{2}\right) \log^2\frac{m^2}{ \mu_{0F}^2}-\left(-\frac{3}{2}+\frac{8 \gamma }{9}+\frac{4 \gamma ^2}{3}\right)
   \log\frac{m^2}{ \mu_{0F}^2}+\frac{2 \zeta (3)}{3}-\frac{\pi ^2}{3}-\frac{56 \gamma }{27}+\frac{3139}{648}\Bigg]
   \nonumber\\
   &+C_F T_R n_\ell
   \Bigg[\left(\frac{2 \gamma }{3}-\frac{1}{2}\right) \log^2\frac{m^2}{ \mu_{0F}^2}-\left(-\frac{3}{2}+\frac{8 \gamma }{9}+\frac{4 \gamma ^2}{3}\right)
   \log\frac{m^2}{ \mu_{0F}^2}
   \nonumber\\
   &\qquad
   -\frac{2 \zeta (3)}{9}+\gamma  \left(\frac{4 \pi ^2}{9}-\frac{4}{27}\right)-\frac{4 \pi ^2}{27}+\frac{8 \gamma ^3}{9}+\frac{8 \gamma
   ^2}{9}-\frac{173}{72}\Bigg].
\end{align}

%% file: app-finite.tex
\section{Analitically-continued Mellin transforms for the $\mathcal{O} \left(\as^2\right)$ initial conditions}
\label{app:mellin}
In this Appendix, we report on the computation of the analitically-continued Mellin transforms of 
the $\mathcal O (\as^2)$ (NNLO) initial conditions, for all fragmenting partons. These are necessary for the inversion of the expressions obtained in Mellin space back to $z$ space. In our case, such
an inversion is performed using the Talbot-path method~\cite{doi:10.1002/nme.995}. 
We remind
the reader that the NNLO initial conditions are taken from~\cite{Melnikov:2004bm,Mitov:2004du}, where they are reported in $z$ space.
Analitically-continued Mellin transforms for most terms can be 
computed e.g. by employing the expressions tabulated in~\cite{Blumlein:1997vf,Blumlein:1998if, Blumlein:2000hw,Blumlein:2003gb,Blumlein:2009ta}. However,
other terms exist for which the analitically-continued Mellin transform cannot be found in literature. For these terms we will
provide details in Sects.~\ref{app:melllog1pz}-\ref{app:mellli3}. Finally, in Sect.\ref{app:mellvalidation}, 
we discuss the validation of our results.

\subsection{Terms including a factor $\frac{\log(1+z)^k}{1+z}$}
\label{app:melllog1pz}

For those terms which contain the factor $\log(1+z)^k/(1+z)$  we apply the expansion
shown in Eqs.~(52),~(56) of~\cite{Blumlein:2000hw}, 
either with an in-house implementation or by using the {\sc ancont} 
code~\cite{Blumlein:1998if, Blumlein:2000hw}.

For example, we have obtained
\begin{eqnarray}
    \mell{\frac{\log(1+z) \log(z)}{1+z}}(N)&=& -\sum_k a_k^{(2)} \frac{k}{2(N-1+k)^2}\,,\\ 
    \mell{\frac{\log^2(1+z) \log(z)}{1+z}}(N)&=& - \sum_k a_k^{(3)} \frac{k}{3(N-1+k)^2}\,,\\ 
    \mell{\frac{\log(1+z) \log^2(z)}{1+z}}(N)&=& \sum_k a_k^{(2)} \frac{k}{(N-1+k)^3}\,,\\ 
    \mell{\frac{\li{2}(-z) \log(z)}{1+z}}(N)&=& -\sum_k a_k^{(2)} \mell{\log(z)}(N-1+k) + \nonumber\\ 
    &&\sum_k a_k^{(1)}\left\{ \mell{\li{2}(-z)}(N-1+k) + \right.\nonumber\\
    &&\left.(N-1)  \mell{\li{2}(-z)\log(z)}(N-1+k)  \right\}\,,
\end{eqnarray}
where the $a_k^{(p)}$ coefficients are tabulated in Ref.~\cite{Blumlein:2000hw}.

We point out that  Eq.~(62) of~\cite{Blumlein:2000hw} has some typographical errors. In the 
convention of that paper (note the exponent $N$ in the
Mellin transform), it should read:
\begin{equation}
    \int_0^1 dz\, z^N \frac{\log(z) \li{2}(z)}{1+z} = -\sum_{k=1}^9 \frac{a_k^{(1)} k}{(N+k)^2} \left[\zeta(2)+\psi'(N+k+1)-2\frac{S_1(N+k)}{N+k}\right]\,,
\end{equation}
where $S_1(N)$ is the first-order harmonic sum~\cite{Vermaseren:1998uu}.

\subsection{Terms including a factor $\frac{1}{(1+z)^p}$}

Since in the results of Refs.~\cite{Melnikov:2004bm,Mitov:2004du} one finds terms containing $f(x) / (1+x)^3$ and $f(x) / (1+x)^4$, we generalise 
Eqs.~(52),~(56) of Ref.~\cite{Blumlein:2000hw} to these cases, and quote the relevant coefficients. Specifically, assuming $p > 1$:
\begin{eqnarray}
    \mell{\frac{f(z)}{(1+z)^{p+1}}}(N) &=& \int_0^1 dz\, z^{N-1} \frac{f(z)}{(1+z)^{p+1}} \nonumber\\
    &=& -\frac{f(1)}{p \,2^{p}} + \frac{1}{p} \int_0^1  dz\, z^{N-2} \frac{(N-1)f(z) + z f'(z)}{(1+z)^{p}}\nonumber\\
    &=&  -\frac{f(1)}{p \,2^{p}} + 
    \frac{1}{p} \sum_k b_k^{(p)}\left\{ (N-1)\mell{f(z)}(N-2+k) + \right.\nonumber\\
    &&\left. \mell{f'(z)}(N-1+k) \right\}\,,
\end{eqnarray}
where we have expanded
\begin{equation}
    \frac{1}{(1+z)^{p}} = \sum_k z^k b_k^{(p)}\,.
    \label{eq:bfdef}
\end{equation}
Relevant cases appearing in the $\mathcal O (\as^2)$ initial conditions are e.g.:
\begin{equation}
    \mell{\frac{\li{2}(z)}{(1+z)^3}},\quad
    \mell{\frac{\log(z)\log(1-z)}{(1+z)^3}}\,.
\end{equation}
In Tab.~\ref{tab:bpcoeffs} we quote the coefficients $b_k^{(p)}$ for $p=1,2,3,4$.
\begin{landscape}
\begin{table}
    \centering
    \begin{tabular}{c|c|c|c|c}
           $k$  &  $b_k^{(1)}$  &  $b_k^{(2)}$  &   $b_k^{(3)}$  &   $b_k^{(4)}$ \\\hline
         0   &   $9.99999964541345\; 10^{-1}$   &   $9.99999933677464\; 10^{-1}$   &   $9.99999982104303\; 10^{-1}$   &   $9.99999982154996\; 10^{-1}$\\
         1   &  $-9.99996402418476\; 10^{-1}$   &  $-1.999992410641282$                &  $-2.999997506715648$   &   $-3.999997298925117$\\
         2   &   $9.99891207372326\; 10^{-1}$   &   $2.999734175752133$                &   $5.999888215742826$   &   $9.99986520926764$\\
         3   &  $-9.98482718182365\; 10^{-1}$   &  $-3.995649728045883$                &  $-9.99760152693402$   &   $-1.999674114142017\; 10^1$\\
         4   &   $9.88168988987737\; 10^{-1}$   &   $4.95979560171482$                 &   $1.497046492308849\; 10^1$   &   $3.495441199631072\; 10^1$\\
         5   &  $-9.42609738737009\; 10^{-1}$   &  $-5.76657557717507$                 &  $-2.076835551058147\; 10^1$   &   $-5.559108874080811\; 10^1$\\
         6   &   $8.14667457989429\; 10^{-1}$   &   $6.08863382410509$                 &   $2.676271285109963\; 10^1$   &   $8.14863786368924\; 10^1$\\
         7   &  $-5.810708426263057\; 10^{-1}$  &  $-5.486499348711234$                &  $-3.12825339595229\; 10^1$   &   $-1.08903801097778\; 10^2$\\
         8   &   $3.048576907895213\; 10^{-1}$  &   $3.889496091354519$                &   $3.168143595654796\; 10^1$   &   $1.285397301817424\; 10^2$\\
         9   &  $-1.009217356923942\; 10^{-1}$  &  $-1.968052398321446$                &  $-2.62122871535319\; 10^1$   &   $-1.280160378039499\; 10^2$\\
        10   &   $1.549613450532657\; 10^{-2}$  &   $6.188462507514354\; 10^{-1}$  &   $1.657366256524406\; 10^1$   &   $1.020968689051037\; 10^2$\\
        11   &   $0$   &   $-8.97364237825404\; 10^{-2}$   &   $-7.377840486156504$   &   $-6.148201198476365\; 10^1$\\
        12   &   $0$   &   $0$   &   $2.035251433168524$   &   $2.591935272452217
  \; 10^1$\\

        13   &   $0$   &   $0$   &   $-2.597997849838223\; 10^{-1}$   &   $-6.761345214073492$\\
        14   &   $0$   &   $0$   &   $0$   &    $8.16915646756799\; 10^{-1}$\\
    \end{tabular}
    \caption{\label{tab:bpcoeffs} Coefficients $ b_k^{(p)}$, defined in Eq.~\protect\ref{eq:bfdef}.}
\end{table}
\end{landscape}
%

\subsection{Terms with the Nielsen functions $\mathbf S_{1,2}(-z)$ and $\mathbf S_{1,2}(z^2)$}
The function $\mathbf S_{1,2}(z)$ is defined as
\begin{equation}
    \mathbf S_{1,2}(z) = \frac{1}{2} \int_0^z \frac{\log^2(1-t)}{t} dt\,,
\end{equation}
from which it trivially follows that 
\begin{equation}
    \mathbf S'_{1,2}(z) =\frac{d  S_{1,2}(z)}{dz} =  \frac{1}{2} \frac{\log^2(1-z)}{z} .
\end{equation}
Given the relation between the Mellin transform of a function and of its derivative,
\begin{equation}
    \mell{f(z)}(N) = \frac{f(1)-\mell{f'(z)}(N+1)}{N}\,,
\end{equation}
and the values
\begin{equation}
    \mathbf S_{1,2}(1)=\zeta(3),\qquad \mathbf S_{1,2}(-1)=\frac{\zeta(3)}{8}\,,
\end{equation}
we then have
\begin{eqnarray}
    \mell{ \mathbf S_{1,2}(z)} &=& \frac{\zeta(3) - \frac{1}{2}\mell{\log^2(1-z)}(N)}{N}\,,\\
    \mell{ \mathbf S_{1,2}(-z)} &=& \frac{\frac{\zeta(3)}{8} - \frac{1}{2}\mell{\log^2(1+z)}(N)}{N}\,.
\end{eqnarray}
By using
\begin{equation}
    \mell{\log^2(1-z)}(N) =\frac{S_1^2(N) + S_2(N)}{N}\,,
\end{equation}
the first equation reads 
\begin{equation}
    \mell{ \mathbf S_{1,2}(z)} = \frac{\zeta(3)}{N} - \frac{1}{2}\frac{S_1^2(N) + S_2(N)}{N^2} \,,
\end{equation}
which agrees with Ref.~\cite{Blumlein:1998if}. 
For $\mathbf S_{1,2}(-z)$, we have instead performed the Mellin transform of $\log^2(1+z)$ using the
expansion shown in Ref.~\cite{Blumlein:2000hw}.

Finally, concerning the Mellin transform of $\mathbf S_{1,2}(z^2)$, we have exploited the relation
\begin{equation}
    \mell{f(x^2)}(N) = \frac{1}{2} \mell{f(x)}\left(\frac{N}{2}\right)\,.
\end{equation}
to rewrite $\mell{\mathbf S_{1,2}(z^2)}$ in terms of $\mell{\mathbf S_{1,2}(z)}$.

\subsection{Terms involving functions of $|2z-1|$}
\label{app:mellabs}
The functions
\begin{eqnarray}
    A_1(z)&=&|2z-1|,\\
    A_2(z)&=&\textrm{ArcTanh}\left(|2z-1|\right)
\end{eqnarray}
appear in the initial conditions of Refs.~\cite{Melnikov:2004bm,Mitov:2004du}, specifically in $d_g^{(2)}$. The Mellin transform of $A_1$ can be easily computed:
 \begin{equation}
     \mell{A_1(z)}(N) = \frac{(N-1)}{N(N+1)}+2\frac{2^{-N}}{N(N+1)}.
     \label{eq:mella1}
 \end{equation}
 Despite its rather simple form, a problem arises when the inverse Mellin transform is considered,
 \begin{equation}
     \int_{c-i \infty}^{c+i\infty} dN \mell{A_1(z)}(N)\, z^{-N}=
     \int_{c-i \infty}^{c+i\infty} dN \frac{(N-1)}{N(N+1)}\, z^{-N}+
     2 \int_{c-i \infty}^{c+i\infty} dN \frac{(2z)^{-N}}{N(N+1)}\,.
     \label{eq:invmellabs}
 \end{equation}
 From Eq.~\eqref{eq:invmellabs}, in particular the second term on the r.h.s., one immediately sees 
 that the integration contour cannot be closed on the left part of the complex plane
 if $z> 1/2$, rather it must be closed on the right, only for this specific term. This poses a practical problem when the inversion  
is performed numerically, as in our case. 
However, since the constant $c$ must be larger than the real part of the left-most pole of the integrand 
 (in this case $c >0$), when the integration contour is closed on the right it does not encircle any pole, hence its 
 contribution is zero. Therefore, a practical solution is to keep the second term on the r.h.s. of
Eq.~\eqref{eq:invmellabs} only when $z<1/2$.\\

Moving on to $A_2$, its Mellin transform can be obtained semi-analytically.
First, while both $A_1$ and $A_2$ have a discontinuous derivative at $z=1/2$, their difference,
\begin{equation}
    A_{\rm diff}(z) = A_2(z)-A_1(z)\,,
\end{equation}
is smooth for $z\in (0,1)$, while at the endpoints it has logarithmic singularities. Subtracting the singularities 
(and imposing that the remainder vanishes for $z=1/2$), one gets the function
\begin{equation}
    A_{\rm reg}(z) = A_{\rm diff}(z) + \frac{1}{2}\left(\log(z) +\log(1-z)\right) +\log(2)\,,
\end{equation}
which is now smooth $z\in [0,1]$, is symmetric around $z=1/2$, and it is normalised such that $A_{\rm reg}(1/2)=0$. $A_{\rm reg}(z)$
can now be fitted with the functional form 
\begin{equation}
    A_{\rm reg}(z)=\sum_{k=0}^{k_{\rm max}} c_k \left(z(1-z)\right)^{k},
    \label{eq:Pfitabs}
\end{equation}
The values of the coefficient $c_k$, obtained with $k_{\rm max} = 7$, are reported in Tab.~\ref{tab:cdekcoeffs}.
In this way the Mellin transform of each term entering the sum can be expressed in terms of the Euler Beta function:
\begin{equation}
    \int_0 ^1 dz z^{N-1+p} (1-z)^p = \mathrm{B}(N+p, p+1)\,.
\end{equation}
To summarise, we rewrite the function $A_2(z)$ in the form
\begin{equation}
    A_2(z)=A_1(z) - \frac{1}{2}\left(\log(z) +\log(1-z)\right) -\log(2) +  \sum_{k=0}^{k_{\rm max}} c_k \left(z (1-z)\right)^k\,,
\end{equation}
where the Mellin transform for all terms on the r.h.s. can now be easily computed.

\begin{table}
    \centering
    \begin{tabular}{c|c|c|c}
          $k$ & $c_k$                               & $d_k$                             & $e_k$                            \\\hline
           0  &  $-3.0691110144628014 \; 10^{-1}$   &  $0$                              &  $-7.089736955408198 \; 10^{-1}$   \\
           1  &  $1.0203235756821032 $              &  $3.6939138307415825 \; 10^{-2}$  &  $2.404090533048734 $              \\
           2  &  $-8.198982139352419 \; 10^{-1}$    &  $6.1086779903441055$             &  $7.919971327024875 $              \\
           3  &  $3.411456568255282 \; 10^1$        &  $-3.17971561004379 \; 10^1  $    &  $-8.829898774056112 \; 10^1$      \\
           4  &  $-4.100490025841941 \; 10^2$       &  $ 1.6434190618411395  \; 10^2$   &  $6.476757783580479 \; 10^2$       \\
           5  &  $2.6287640325082357 \; 10^3$       &  $-6.293777861195324  \; 10^2$    &  $-2.849867107259887 \; 10^3$      \\
           6  &  $-8.382883186201163 \; 10^3$       &  $ 1.6880992305561952  \; 10^3$   &  $6.756475895565204 \; 10^3$       \\
           7  &  $1.0667992579287005 \; 10^4$       &  $-3.142029544696374  \; 10^3$    &  $-6.616331933715405 \; 10^3$      \\
           8  &  $0$                                &  $ 4.016414152661526   \; 10^3$   & $0$                                \\
           9  &  $0$                                &  $-3.4409243129611673  \; 10^3$   & $0$                                \\
           10 &  $0$                                &  $ 1.8763407749751084 \; 10^3$    & $0$                                \\
           11 &  $0$                                &  $-5.83979122751871   \; 10^2$    & $0$                                \\
           12 &  $0$                                &  $ 7.805619459373399  \; 10^1$    & $0$                                \\
    \end{tabular}
    \caption{\label{tab:cdekcoeffs} Coefficients $c_k$, $d_k$, and $e_k$, 
        defined in Eqs.~\protect\ref{eq:Pfitabs}, \protect\ref{eq:Pfitli2} and \protect\ref{eq:Pfitli3}, respectively.}
\end{table}

\subsection{Terms involving $\li{2}\left(\frac{2z-1}{z}\right)$}
\label{app:mellli2}
We will consider the Mellin transforms of the following expressions:
\begin{eqnarray}
F_1(z)&=&\li{2}\left(\frac{2z-1}{z}\right),\\
F_2(z)&=&\li{2}\left(\frac{2z-1}{z}\right)\log z\,,\\
F_3(z)&=&\li{2}\left(\frac{2z-1}{z}\right)\log (1-z).
\end{eqnarray}
We will focus on the case for $F_1$. This function is divergent at $z\to0$, therefore we proceed like in sec.~\ref{app:mellabs}, 
exposing the singular behaviour in this region. However, at variance with what we did there, the remainder is bounded but its derivative
is still divergent both at $z\to 0$ and $z\to 1$. These terms have also to be subtracted, in order to have a regular function 
${F_1}_{\rm reg}$ to fit. We call ${F_1}_{\rm sing}^0$ the singular part of $F_1$ at $z\to 0$,
and ${F_1}_{d-\rm sing}^0$, ${F_1}_{d-\rm sing}^1$ the contributions which make the derivative of 
$F_1 - {F_1}_{\rm sing}^0$ divergent in $z\to 0$ and $z\to 1$ respectively~\footnote{
    Constants and regular terms may also be included in ${F_1}_{\rm sing}$, ${F_1}_{d-\rm sing}$.}
    . Thus we have
\begin{equation}
    F_1(z) ={F_1}_{\rm sing}^0 (z)+ {F_1}_{d-\rm sing}^0 (z)+ {F_1}_{d-\rm sing}^1(z) + {F_1}_{\rm reg} (z)\,,
\end{equation}
where
\begin{eqnarray}
    {F_1}_{\rm sing}^0 (z) &=&-\frac{\pi^2}{6} -\frac{1}{2} \log^2 z \,,\label{eq:f1sing0}\\ 
    {F_1}_{d-\rm sing}^0 (z) &=& z - 2z \log z\,,\label{eq:f1dsing0}\\
    {F_1}_{d-\rm sing}^1 (z) &=& z + (1-z) \log (1-z)\,.\label{eq:f1dsing1}
\end{eqnarray}
The remainder ${F_1}_{\rm reg}$, which we have chosen such that ${F_1}_{\rm reg}(0)=0$, can now be fit with a polinomial in $z$
\begin{equation}
     {F_1}_{\rm reg}(z) = \sum_{k=1}^{k_{\rm max}} d_k z^k\,,
     \label{eq:Pfitli2}
\end{equation}
where we have used $k_{\rm max}=12$. The coefficients $d_k$ are shown in tab.~\ref{tab:cdekcoeffs}.

For what concerns the functions $F_2$ and $F_3$, the terms in Eqs.~\ref{eq:f1sing0}-\ref{eq:f1dsing1} are simple enough
so that the Mellin transform of their product with either $\log z$ or $\log(1-z)$ can be computed with
standard methods. The Mellin transform of the corresponding regular part is also trivial, since it amounts just to the
Mellin transform of $\log z$ or $\log(1-z)$ multiplied by a power of $z$.

\subsection{Mellin transform of $\li{3}\left(\frac{2z-1}{z}\right) + \li{3}\left(\frac{2z-1}{1-z}\right)$ }
\label{app:mellli3}
The combination
\begin{equation}
G(z) = \li{3}\left(\frac{2z-1}{z}\right) + \li{3}\left(\frac{2z-1}{1-z}\right)
\end{equation}
appears in the gluon initial condition. Exploiting its symmetry around $z=1/2$, we can proceed
in a similar way as in Sects.~\ref{app:mellabs} and~\ref{app:mellli2}. The function $G$ is smooth for $z\in (0,1)$, thus we proceed
by extracting the singular parts of the function, and of the derivative of the remainder, when $z\to 0$ and $z\to 1$:
\begin{equation}
    G(z) = G_{\rm sing}^0 (z)+ G_{\rm sing}^1(z) + G_{d-\rm sing}^0 (z)+ G_{d-\rm sing}^1(z) + G_{\rm reg} (z)\,,
\end{equation}
 The singular parts,
written so that the remainder satisfies s $G_{\rm reg}(1/2) = 0$, read
\begin{eqnarray}
    G_{\rm sing}^0 (z) &=& \frac{\log(z)}{6}\left[  \pi^2 + \log^2(z)\right] 
                + L_2\left(\frac{\pi^2}{6} -L_2 + \frac{L_2^2}{6} \right)\,,\\
    G_{d-\rm sing}^0 (z) &=& z \log^2(z) \,,\\
    G_{\rm sing}^1 (z) &=& G_{\rm sing}^0 (1-z)\,,\\
    G_{d-\rm sing}^1 (z) &=&  G_{d-\rm sing}^0 (1-z)\,,
\end{eqnarray}
where $L_2=\log(2)$. Following what we did in Sect.~\ref{app:mellabs} (see Eq.~\eqref{eq:Pfitabs}), we
fit the regular part $G_{\rm reg} (z)$ with the functional form
\begin{equation}
    A_{\rm reg}(z)=\sum_{k=0}^{k_{\rm max}} e_k \left(z(1-z)\right)^{k},
    \label{eq:Pfitli3}
\end{equation}
where $k_{\rm max}=7$ and the coefficients $e_k$ are given in Tab.~\ref{tab:cdekcoeffs}.

\subsection{Validation}
\label{app:mellvalidation}
\begin{figure}[ht!]
    \includegraphics[width=0.49\textwidth]{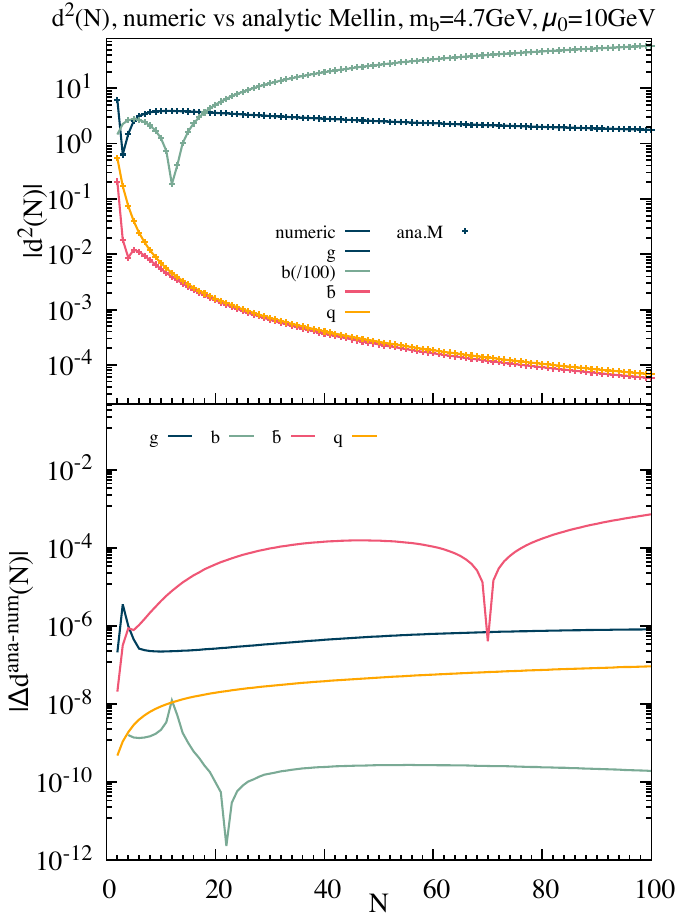}
    \includegraphics[width=0.49\textwidth]{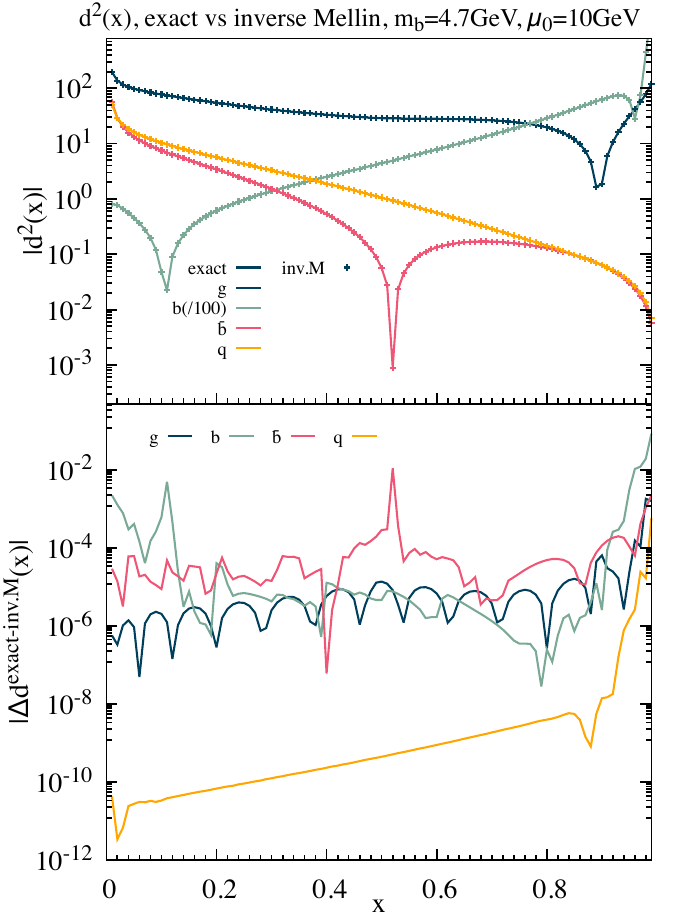}
    \caption{\label{fig:valMell} Validation plots in $N$ (left) and $z$ space for the Mellin transforms of the $\mathcal O(\as^2)$
initial conditions. See the text for details.}
\end{figure}
We conclude this Appendix by presenting some validation plots for the analytically-continued Mellin transforms of the $\mathcal O(\as^2)$
initial conditions. For all the cases presented in this Appendix, 
the fitting parameters have been chosen so that the fitted function differs from the exact one by no more than few parts in $10^5$.
Our validation is both in $N$ and in $z$ space, setting $\mu_0\ne m$ (specifically, $m=4.7\,\gev$, $\mu_0=10\,\gev$) in order to expose all terms 
of the initial conditions. Starting from $N$ space, we compare in the left plot of Fig.~\ref{fig:valMell} 
results obtained by computing numerically 
the Mellin transform of the expressions in Refs.~\cite{Melnikov:2004bm,Mitov:2004du}, by employing a Gaussian integrator, 
with the analytic Mellin transforms obtained as described above.
In the upper panel of that figure, we show as lines (symbols) the absolute value of the numeric (analytic) Mellin transforms
of the initial conditions. In the lower panel, we plot the quantity
\begin{equation}
    \Delta d_X ^{\rm ana-num} = \left|\frac{d_X^{\rm ana} - d_X^{\rm num}}{d_X^{\rm ana} + d_X^{\rm num}}\right|\,,
    \label{eq:deltaD}
\end{equation}
where $X=b, \overline b , g, q$, 
i.e. the relative difference between the numeric and analytic Mellin transforms. From this panel we appreciate that
the difference between the numeric and analytic Mellin transforms never exceeds $10^{-4}$, with the exception
of $D_{\overline b}$ at very large $N$.

The validation in $z$ space, shown in the right plot of Fig.~\ref{fig:valMell}, presents a very similar layout,
with the only relevant change being that the numeric (analytic) Mellin transforms of the initial conditions
are replaced with their exact expression (inverse Mellin transform) in $x$ space. Again, the relative difference does not
exceed $10^{-4}$, except for those points where the initial conditions, hence the denominator in Eq.~\eqref{eq:deltaD}, vanish, or
in the region $z\to1$. Here it has to be stressed that we do not plot the contribution coming from Dirac delta's 
at $z=1$ or the endpoint of plus distributions, which live exactly at $z=1$, which we deem responsible for such a discrepancy. Indeed, we 
have observed similar discrepancies also in the case of very simple distributions, e.g. $\left[\frac{\log(1-z)}{1-z}\right]_+$.